\documentclass[12pt]{article}
\usepackage{amsmath}
\usepackage{amsfonts}
\usepackage{amssymb}
\usepackage{graphicx}

\newif{\ifcomentarios}
\comentariosfalse

\input{tcilatex}

\begin{document}

\title{A hopping mechanism for cargo transport by molecular motors in
crowded microtubules. }
\author{Carla Goldman \\
Departamento de F\'{\i}sica Geral - Instituto de F\'{\i}sica \\
Universidade de S\~{a}o Paulo CP 66318\\
05315-970 S\~{a}o Paulo, Brazil.}
\date{May 2010}
\maketitle

\begin{abstract}
Most models designed to study the bidirectional movement of cargos as they
are driven by molecular motors rely on the idea that motors of different
polarities can be coordinated by external agents if arranged into a
motor-cargo complex to perform the necessary work \cite{gross04}. Although
these models have provided us with important insights into these phenomena,
there are still many unanswered questions regarding the mechanisms through
which the movement of the complex takes place on crowded microtubules. For
example (i) how does cargo-binding affect motor motility? and in connection
with that - (ii) how does the presence of other motors (and also other
cargos) on the microtubule affect the motility of the motor-cargo complex?
We discuss these questions from a different perspective. The movement of a
cargo is conceived here as a \textit{hopping process} resulting from the
transference of cargo between neighboring motors. In the light of this, we
examine the conditions under which cargo might display bidirectional
movement even if directed by motors of a single polarity. The global
properties of the model in the long-time regime are obtained by mapping the
dynamics of the collection of interacting motors and cargos into an
asymmetric simple exclusion process (ASEP) which can be resolved using the
matrix \textit{ansatz} introduced by Derrida \cite{derrida}.

\vspace{0.2in}\vspace{0.25in}

\textit{keywords -} intracellular transport by molecular motors;
bidirectional movement of cargo, traffic jam on microtubules; ASEP models.
\end{abstract}

\section{\protect\bigskip Introduction}

Research interest in the origins of the long-range bidirectional movement of
particles (organelles, vesicles, nutrients) driven by molecular motors is
motivated by fundamental questions concerning the nature of interactions
between motors and their cargos as transport processes take place. A current
explanation for the phenomenon relies on the idea that motors of different
polarities act coordinately on the same particle at different times. If,
however, they act in parallel, the bidirectional movement would reflect
dominance of one or another kind of motor achieved by a \textit{tug-of-war}
mechanism \cite{gross04}, \cite{welte04}, \cite{zeldovich}, \cite{mallick04}%
, \cite{klumpp}. An important question that remains in this context concerns
the mechanisms that would promote such coordination \cite{ma}. Alternatives
to the coordination or \textit{tug-of-war} models in the literature arise
from the possibility of attributing the phenomenon to a dynamic role of the
microtubules \cite{kulic} or to a mechanical coupling between different
motors \cite{gelfand 09}.

A general difficulty encountered within any of these views is related to the
presence of other particles (including other motors) on the microtubule at a
given time that are not directly involved with the transfer process. These
other particles are expected to impose restrictions on motility and
performance of the motors that are directly interacting with cargo at that
time \cite{lipowsky}. Contrarily to these expectations, however, data from
observations of beads driven by kinesins in steady-state conditions indicate
that the number of long length runs of such beads increases significantly as
the density of motors at the microtubule increases, although their
velocities remain essentially unaltered within a wide range of motor
concentrations \cite{seitz}, \cite{beeg}. Thus, the reality of traffic jam
in crowded microtubules still challenges the current view of long-range
cargo transport that presupposes an effective and controllable movement\ of
the motor(s) arranged into a motor-cargo complex. This, of course, requires
a certain degree of stability of motor-cargo interactions and motor
processivity.

Our intention here is to discuss these problems from a different perspective
by bringing into this scenario the model introduced in \cite{goldman1} to
examine cargo transport as a \textit{hopping} process. According to that,
motors and cargos would not assemble into complexes to put transport into
effect. On the contrary, each motor would function as an active overpass for
cargo to step over to a neighboring motor. \ In this case, the long-range
movement of cargo is envisaged as a sequence of these elementary
(short-range) steps either forwards or backwards. In \cite{goldman1} we
examined the conditions under which this may happen, accounting for the fact
that motor motility is affected by the interactions with other motors and
with cargos on the microtubule. There, we considered the presence of a
collection of interacting motors, all of them presenting the same polarity
(kinesins may be thought of as prototypes) and a single cargo. Here, we
examine whether it is possible to explain in a similar context the origin of
the observed bidirectional movement displayed by cargos.

The particular mechanism we propose to substantiate the hopping differs from
that suggested in \cite{goldman1}. It keeps, however, the same general ideas
of the original. As it will be explained below, we view the hopping of cargo
between motors as an effect of thermal fluctuations undergone by motor
tails. The flexibility of the tails may promote contact and, eventually,
exchange of cargo between neighboring motors.

As in \cite{goldman1}, the model dynamics is mapped into an asymmetric
simple exclusion process (ASEP) \cite{pablo kipnis}, \cite{derrida lebowitz}%
, \cite{review evans} whose stationary properties are resolved explicitly in
the limit of very large systems. Other ASEP models have already been
considered in the literature to study the conditions for motor jamming in
the absence of cargo \cite{lipowsky}, \cite{parmeggiani}, \cite{chowdhury}.
Our model is conceived to account explicitly for changes in the dynamics of
the motors that at a certain instant of time are interacting with cargos.

The model is reviewed here in order to include a second cargo in the system,
still keeping the presence of motors of a single polarity. We believe that
this approaches more realistic situations in which the simultaneous presence
of many cargos and motors on the same microtubule must be the prevailing
situation \cite{kulic}. We show that under these conditions, a cargo may be
able to execute long-range bidirectional movement as it moves over clusters
of motors assembled either at its back end or at the back end of the cargo
in front. One may recognize in this a possibility for explaining the origins
of self-regulation in intracellular transport since it has been suggested in
the last few years that signaling pathways involved in intracellular traffic
regulation can be\ performed simply by the presence of cargos at the
microtubule \cite{verhey}. We then speculate that the passage of cargos on
microtubules does not get blocked by motor jamming. On the contrary, jamming
operates as an allied process to promote long runs of cargos across motor
clusters. In this case, the density of motors on the microtubule can be
identified as an element of control in intracellular transport since it
directly affects the conditions for jamming.

It is worth mentioning that the model developed here does not rule out other
possibilities, such as the \textit{tug-of-war} or competition models. What
we suggest is that the presence of motors of different polarities may not be
essential to explain the origin of the bidirectional movement.

The hopping mechanism is presented in Sec.2. The kinetic properties of the
extended version are developed in Sec.3, considering the presence of two
cargos. In Sec.4 we present our results. Additional remarks and conclusions
are in Sec.5.

\section{A{}n alternative for cargo driving}

The stochastic model in \cite{goldman1} formulated in a lattice describes
the dynamics of motors and cargos accounting for (i) steric interactions
among different particles moving on the same microtubule; (ii) the presence
of motors of a single polarity; (iii) the fact that cargos do not move if
not driven by motors.

The crucial point is in item (iii) because it requires a specific model for
motor-cargo dynamics as transportation takes place. We offer here a slightly
different view from that in \cite{goldman1} keeping however the reliance on
the ability of motors to transfer cargo. One way by which this may be
achieved is sketched in \emph{Figure (1)}. The stepping of cargo would be
accomplished as it is released from a motor to which it is attached at a
certain instant of time and then get attached to another (neighboring) motor
either at the left or at the right -- see \emph{Figure (1a)}. This process,
like the one discussed in \cite{goldman1}, relies strongly on the
flexibility of the motor%
\'{}%
s tail. The idea was inspired by experimental results suggesting a dynamic
role of the kinesin%
\'{}%
s coiled-coil segment in the process \cite{imanishi}, \cite{tomishigue} and
also by data indicating that under load, kinesin motors display an
oscillatory movement \cite{nishiyama} \cite{yanagida}. Here, we think of
these oscillations as signaling fluctuations in the position of motor%
\'{}%
s tail, not necessarily being correlated with displacement of its center of
mass. If this is the case, such oscillations would promote contact between
neighboring motors favoring cargo exchange.

Accordingly, long-range displacements of cargo would reflect a hopping
process extended over many neighboring motors which may be accomplished if
these motors get jammed into clusters for sufficient long periods of time.\
Notice that the whole mechanism does not require special stability of
cargo-motor binding. On the contrary, the transfer of cargo by a motor would
be ease by a loose attachment between them.

The model in \cite{goldman1} was resolved explicitly considering the
presence of a single cargo in the system. The averages for the quantities of
interest were determined in steady-state conditions. We showed there that
the long-range displacements of the cargo would occur predominantly in the
backwards direction, i.e. in opposition to the direction of the movement of
the considered motors. We shall show here that the same dynamics may lead
cargo to display bidirectional movement if the system contains at least one
more cargo interfering with the movement of the motors.

To examine the properties of the system containing two cargos and an
arbitrary number of motors, we map it into the same ASEP as in \cite%
{goldman1} whose dynamics can also reproduce the scheme in \emph{Figure(1)}.

\subsection{Movement of motors and cargos: the ASEP model with two cargos}

We consider a one-dimensional lattice with $M$ sites, representing the
microtubule with periodic boundary conditions. This system contains $N$
motors and a number $K$ of other particles - the cargos - that interact with
motors in order to move. Each site can be occupied by a motor or by a motor
attached to a cargo \emph{(see Fig.1(a))}, otherwise it is empty. The total
number of sites that remain unoccupied is $\ G=M-N-K>0$. Here, we analyze
the long-time behavior of this system for $K=2$ and determine the average
cargo velocity as a function of the parameters. The results indicate
conditions for cargos to perform a type of long-range movement that share
the characteristics of the observed bidirectional movement.

The map of the dynamics shown in \emph{Figure.1} into the considered ASEP is
carried out as follows.\emph{\ }First, each site is identified by its
position $j=1,2,...N$ at the lattice. Then, to each of these sites is
associated a variable $\sigma _{j}$ that assumes integer values $0,1$ or $2$
such that $\sigma _{j}=0$ if the site $j$ is empty, $\sigma _{j}=1$ if it is
occupied by a motor; or $\sigma _{j}=2$ if it is occupied by a motor
attached to a cargo. With these, a configuration $C$ of the lattice is
specified by the set $\{\sigma _{1}\sigma _{2}...$ $\sigma _{N}\}.$ The
dynamics of the ASEP that reproduces the elementary steps in \emph{Fig.1}
can now be defined. For this, consider that at each time interval $dt$ a
pair of consecutive sites, say $j$ and $j+1$ are selected at random. The
occupancy of these two sites is then switched according to the following
rules 
\begin{equation}
\begin{array}{cccccc}
\mathbf{(a)} & 10 & \rightarrow & 01 & \text{with rate }k, & \hspace{0.1in}%
\text{probability }kdt \\ 
\mathbf{(b)} & 12 & \rightarrow & 21 & \text{with rate }w, & \hspace{0.1in}%
\text{probability }wdt \\ 
\mathbf{(c)} & 21 & \rightarrow & 12 & \text{with rate }p, & \hspace{0.1in}%
\text{probability }pdt%
\end{array}
\label{dinamica}
\end{equation}%
where the pair $(j,j+1)$ is represented by the values of the corresponding
site variables $(\sigma _{j},\sigma _{j+1}).$The parameters $k,w$ and $p$
are the assigned probabilities per unit time (rates) for occurrence of the
processes indicated. Process $(a)$ describes the possibility for a motor
(kinesin) that carries no cargo to step forward to a neighboring empty site 
\emph{(Figure 1b)}. Processes $(b)$ and $(c)$ account for the switching of
the cargo between two neighboring motors. This accounts either for backward $%
(b)$ or for forward steps $(c)$. Notice that the dynamics conserves the
number of particles of type-1 as well as those of type-2.

In order to investigate the long-time dynamics of a cargo resulting from
these elementary steps we use the \textit{matrix} \textit{ansatz} introduced
by Derrida \cite{derrida}, \cite{review evans}. The idea is to represent the
probability $P_{N,M}(C)$ of a configuration $C$ of the system with $N$ sites
and $M$ particles of type-1 as a trace over a product of $N$ non-commuting
matrices, each specifying the corresponding site occupancy:%
\begin{equation}
P_{N,M}(C)=\dfrac{1}{Z_{N,M}}Tr\displaystyle\prod\limits_{i=1}^{N}(\delta
_{\sigma _{i},1}D+\delta _{\sigma _{i},2}A+\delta _{\sigma _{i},0}E)
\label{prob do estado}
\end{equation}%
where 
\begin{equation}
Z_{N,M}=\sum_{\{\sigma _{i}\}}Tr\displaystyle\prod\limits_{i=1}^{N}(\delta
_{\sigma _{i},1}D+\delta _{\sigma _{i},2}A+\delta _{\sigma _{i},0}E)
\label{norma}
\end{equation}%
is the normalization. The sum runs over all configurations for which $%
\sum_{i}^{N}\delta _{\sigma _{i},1}=M$ \ and $\sum_{i}^{N}\delta _{\sigma
_{i},2}=K=2$. In this product, a site $i$ is represented by a matrix $D$ if
it is occupied by a motor ($\sigma _{i}=1$) or by a matrix $A$ if occupied
by a motor with a cargo ($\sigma _{i}=2$); if the site is empty it is
represented by a matrix $E$ ($\sigma _{i}=0$). In order to calculate
averages over these configurations in the stationary state, it is necessary
at first to find the \textit{algebra} that must be satisfied by these
matrices such that the probabilities defined in (\ref{prob do estado})
satisfy the stationary conditions \cite{derrida lebowitz},

\begin{equation}
\sum_{C^{\prime }}P_{N,M}(C^{\prime })\Gamma (C^{\prime }\rightarrow
C)-P_{N,M}(C)\Gamma (C\rightarrow C^{\prime })=0  \label{mestra}
\end{equation}%
where the sum extends over all configurations of $M$ motors distributed over 
$N-K$ lattice sites. Observe that the nonzero terms on the LHS of the above
equation are those for which configurations $C$ and $C^{\prime }$ differ
from each other at most by the positions of a pair of consecutive sites,
which can be reversed by any of the elementary processes defined by the
dynamics in (\ref{dinamica}). In this case, each factor $\Gamma (C^{\prime
}\rightarrow C)$ (or $\Gamma (C\rightarrow C^{\prime })$) must be replaced
by the rate $w$, $k$ or $p$ for the corresponding elementary process that
brings $C$ back from $C^{\prime }$ (or $C^{\prime }$ from $C$ ).

The algebra corresponding to the ASEP defined by the dynamics (\ref{dinamica}%
) has been presented in (\cite{goldman1}) for $K=1:$

\begin{equation}
\begin{array}{c}
DA-xAD=E-D \\ 
DE=E \\ 
EA=E \\ 
EE=E%
\end{array}
\label{algebra}
\end{equation}%
with 
\begin{equation}
x=\frac{k+p}{w}  \label{x}
\end{equation}%
Here, we shall use this same algebra to evaluate the traces over products of
matrices $D,$ $A$ and $E$ that appear in calculating averages over the
quantities that characterize the movement of a cargo. Before proceeding,
however, a few remarks are in order.

\subsection{Traffic profile in the system with two cargos}

The model with two cargos is not ergodic. The dynamics preserves the number
of empty spaces in each of the two partitions defined by the initial
positions of the two cargos in the system with periodic boundary conditions.
\ In this case, all configurations $C$ and $C%
{\acute{}}%
$ that satisfy equation (\ref{mestra}) must share the number of empty spaces
in each of the partitions. Moreover, configurations in which the empty
spaces are all concentrated in one of the two partitions must be excluded,
for these do not satisfy (\ref{mestra}) with the algebra (\ref{algebra}).

We treat the initial conditions (\textit{IC}), namely the number of empty
spaces - $h$ in one of the partitions and $G-h$ in the other, as \textit{%
random variables}. This artifact shall account for the uncertainty one has
in experimental data regarding the relative positions of the particles, and
also for effects of random processes that are not explicitly described by
the present model such as motor binding and unbinding at the microtubule.
For computing averages, we shall account first for all possible
configurations at fixed $h$ and then average the results over $h$. The
procedure is further specified observing that (a) because there is no reason
to favor any initial configuration, we may consider that $h$ is uniformly
distributed and (b) in analogy with a situation of equilibrium, we take the
average \textit{annealing }as the averages over particle configurations are
performed in parallel with average over $h.$

The measure $P_{N,M}(C_{(h)})$ of a configuration $C_{(h)}$ of a \textit{%
subset}-$h$ is written as the trace over a product of matrices $A,D$ and $E$
that satisfy the algebra in (\ref{algebra}):

\begin{eqnarray}
P_{N,M}(C_{(h)}) &=&\dfrac{1}{Z_{N,M}^{\left( h\right) }}%
Tr(D^{p_{1}}E^{k_{1}}D^{p_{2}}E^{k_{2}}...D^{p_{k}}E^{k_{k}}\mathbf{A}%
D^{p_{k+1}}E^{k_{k+1}}...  \label{prob do estado2} \\
&&...D^{p_{p}}E^{k_{p}}\mathbf{A}%
D^{p_{p+1}}E^{k_{p+1}}...D^{p_{N-1}}E^{k_{N-1}}D^{p_{N}}E^{k_{N}})  \notag
\end{eqnarray}%
In the expression above, each $p_{i}$ is a binary variable such that $%
p_{i}\in \{0,1\},$ $i=1,2,...N$ and $k_{i}=1-p_{i}$ satisfying 
\begin{equation}
\begin{array}{c}
k_{1}+k_{2}+...+k_{k}+k_{p+1}+...+\text{ }k_{N-1}+k_{N}=h\text{ } \\ 
\text{and} \\ 
k_{k+2}+k_{k+3}+...+k_{p}=G-h%
\end{array}
\label{vinculo}
\end{equation}

The normalization 
\begin{equation}
Z_{N,M}^{\left( h\right)
}=W_{2-2}^{(h)}=W_{12-12}^{(h)}+W_{02-12}^{(h)}+W_{12-02}^{(h)}+W_{02-02}^{(h)}~\ 
\label{ZNM(h)}
\end{equation}%
is conveniently expressed in terms of the weights $W_{\left( \sigma
_{i-1}\sigma _{i}\sigma _{i+1....}\right) -\left( \sigma _{j-1}\sigma
_{j}\sigma _{j+1....}\right) }^{(h)}$. These are defined as the sum over the
traces corresponding to the configurations that belong to the subset $h$ for
which the occupation of the sites ... $i-1,i,i+1$... and ...$j-1,j,j+1$...
in the \textit{n-tuples }are\textit{\ }fixed and\textit{\ }specified by the
values of the corresponding site variables $(\sigma _{i-1},\sigma
_{i},\sigma _{i+1},...)$ and $(\sigma _{j-1},\sigma _{j},\sigma _{j+1},...),$
respectively.

The $P_{N,M}(C_{(h)})$ defined above must satisfy the stationary conditions 
\begin{equation}
\sum_{C_{(h)}^{\prime }}P_{N,M}(C_{(h)}^{\prime })\Gamma (C_{(h)}^{\prime
}\rightarrow C_{(h)})-P_{N,M}(C_{(h)})\Gamma (C_{(h)}\rightarrow
C_{(h)}^{\prime })=0~~  \label{mestra-h}
\end{equation}%
where the sum extends over all configurations $C_{(h)}^{\prime }$ that
belong to the subset $h$. The transition rates $\Gamma (C_{(h)}^{\prime
}\rightarrow C_{(h)})$ lead configurations $C_{(h)}^{\prime }$ into
configurations $C_{(h)}$.

\subsection{The average velocity of a cargo}

Consistently with the above definitions we represent the average value$\quad
<v_{(h)}>$ \quad of the velocity of any of the two cargos at fixed $h$ as 
\begin{equation}
<v_{(h)}>\text{ }=\frac{1}{Z_{N,M}^{(h)}}\left\{ pW_{\left( 21\right)
-\left( 2\right) }^{(h)}-wW_{\left( 12\right) -\left( 2\right)
}^{(h)}\right\}  \label{vmedio(a)2h}
\end{equation}%
The configurations associated with $W_{\left( 21\right) -\left( 2\right)
}^{(h)}$ are such that the specified cargo has one motor at its right side
that allows it to move one step to the right at a rate $p$. Similarly, in
the configurations associated to $W_{\left( 12\right) -\left( 2\right)
}^{(h)}$ there is a motor at the left side of this cargo that allows it to
move one step to the left at rate $w$. In both types of configurations the
neighborhood of the other cargo is not specified.

It shall be convenient to subdivide the above sum into sums over
configurations having the same trace. This is achieved by specifying in (\ref%
{vmedio(a)2h}) the occupation of the sites that precede both cargos. For
this, $<v_{(h)}>$ is rewritten as 
\begin{eqnarray}
&<&v_{(h)}>\text{ }=\frac{1}{Z_{N,M}^{(h)}}\left\{ p\left[ W_{\left(
121\right) -\left( 12\right) }^{(h)}+W_{\left( 121\right) -\left( 02\right)
}^{(h)}+W_{\left( 021\right) -\left( 12\right) }^{(h)}+W_{\left( 021\right)
-\left( 02\right) }^{(h)}\right] \right.  \notag \\
&&\left. -w\left[ W_{\left( 12\right) -\left( 12\right) }^{(h)}+W_{\left(
12\right) -\left( 02\right) }^{(h)}\right] \right\}  \label{vmedio(b)2h}
\end{eqnarray}

To proceed in the evaluation of (\ref{vmedio(b)2h}) it is also convenient to
replace the site variables $\{\sigma _{i}\}$ by block variables $\{m_{i}\}$
and $\{q_{i}\}$ $i=1,2...k$, that assume integer values to represent,
respectively, sequences of motors and empty sites in a configuration $%
C_{(h)}.$ With these, the sum over configurations that contribute to $%
W_{\left( 121\right) -\left( 12\right) }^{(h)}$ for example, can be
expressed as 
\begin{eqnarray}
W_{\left( 121\right) -\left( 12\right) }^{(h)} &=&\displaystyle%
\sum\limits_{\{q_{i}\}}{}^{\prime }\displaystyle\sum\limits_{\{m_{i}\}}{}^{%
\prime }tr(D^{m_{1}}E^{q_{1}}D^{m_{2}}E^{q_{2}}...  \notag \\
&&...D^{m_{k}}E^{q_{k}}D^{m_{k+1}}AD^{m_{k+2}}E^{q_{k+2}}...D^{m_{p}}E^{q_{p}}D^{m_{p+1}}A)
\label{W121 12 (h)}
\end{eqnarray}%
with $m_{k+1},m_{k+2},m_{p+1}\geq 1$. Here, $D^{m_{i}}$ $($or $E^{q_{i}})$
indicates a product of $m_{i}$ $($or $q_{i})$ matrices $D$ $($or $E)$. The
symbol on the summation signals indicates that these are restricted to the
configurations that satisfy the constraints in (\ref{vinculo}) for $0<h<G$.

All the traces in the RHS of (\ref{vmedio(b)2h}) can now be reduced with the
aid of the algebra in (\ref{algebra}). For this, we use the identity 
\begin{equation}
D^{K}AE=x^{K}AE  \label{identity}
\end{equation}%
that also follows directly from (\ref{algebra}) \cite{goldman1}. The results
are quoted as follows%
\begin{equation}
\begin{array}{cll}
W_{\left( 12\right) -\left( 12\right) }^{(h)} & =\displaystyle%
\sum\limits_{conf}{}^{\prime
}tr(ED^{m_{i}}AED^{m_{k}}ED^{m_{j}}AED^{m_{k+1}}) & =\displaystyle%
\sum\limits_{conf}{}^{\prime }x^{m_{i}}x^{m_{j}}tr(E) \\ 
W_{\left( 121\right) -\left( 12\right) }^{(h)} & =\displaystyle%
\sum\limits_{conf}{}^{\prime }tr(ED^{m_{i}}AD^{m_{k}}ED^{m_{j}}A) & =%
\displaystyle\sum\limits_{conf}{}^{\prime }x^{m_{i}}x^{m_{j}}tr(E) \\ 
W_{\left( 12\right) -\left( 02\right) }^{(h)} & =\displaystyle%
\sum\limits_{conf}{}^{\prime }tr(ED^{m_{i}}AED^{m_{j}}EA) & =\displaystyle%
\sum\limits_{conf}{}^{\prime }x^{m_{i}}tr(E) \\ 
W_{\left( 021\right) -\left( 12\right) }^{(h)} & =\displaystyle%
\sum\limits_{conf}{}^{\prime }tr(EAD^{m_{i}}ED^{m_{j}}A) & =\displaystyle%
\sum\limits_{conf}{}^{\prime }x^{m_{i}}tr(E) \\ 
W_{\left( 121\right) -\left( 02\right) }^{(h)} & =\displaystyle%
\sum\limits_{conf}{}^{\prime }tr(ED^{m_{i}}AD^{m_{j}}EA) & =\displaystyle%
\sum\limits_{conf}{}^{\prime }x^{m_{i}}tr(E) \\ 
W_{\left( 021\right) -\left( 12\right) }^{(h)} & =\displaystyle%
\sum\limits_{conf}{}^{\prime }tr(EAD^{m_{i}}ED^{m_{j}}A) & =\displaystyle%
\sum\limits_{conf}{}^{\prime }x^{m_{i}}tr(E) \\ 
W_{\left( 02\right) -\left( 02\right) }^{(h)} & =\displaystyle%
\sum\limits_{conf}{}^{\prime }tr(EAEA) & =\displaystyle\sum%
\limits_{conf}{}^{\prime }tr(E)%
\end{array}
\label{tracos Tabela1}
\end{equation}

The above expressions are independent of $h$. The only dependence on $h$ in
the evaluation of the weights comes from the multiplicity of the
configurations. Configurations for which $h$\textbf{\ }$=0$\textbf{\ or }$%
h=G $\textbf{\ }contribute with a factor $1/N$ with respect to the
contributions from all other configurations that result in the same trace.

\subsection{Average over the random variable}

We now take the average of $<v_{(h)}>$ over all realizations of $h$ that
assumes an integer value within the interval $h\in \lbrack 1,G-1]$ for $G>2$%
, with equal probability. This is performed here as%
\begin{equation}
<v>\smallskip =\frac{\sum_{h}\{pW_{\left( 21\right) -\left( 2\right)
}-wW_{\left( 12\right) -\left( 2\right) }\}}{\sum_{h}Z_{N,<M>}}\equiv \frac{%
\{p\overline{W}_{\left( 21\right) -\left( 2\right) }-w\overline{W}_{\left(
12\right) -\left( 2\right) }\}}{\overline{Z}_{N,<M>}}  \label{<V(2)>}
\end{equation}%
that corresponds to the average \textit{annealing} in analogy to a situation
of equilibrium. The averaged quantities are indicated by the bars over the
corresponding symbols representing the weights $\overline{W}$ and
normalization $\overline{Z}$.

Now, observe that because the traces do not depend on $h$, then the sums in
the above expression, both in the numerator and in the denominator, account
for all possible configurations of arbitrary sequences of empty and occupied
sites, keeping fixed just the \textit{n-tuples} indicated in each term. With
this, the restrictions imposed on the sums in (\ref{tracos Tabela1}) are
removed.

\subsection{Sum over configurations}

We estimate the number of configurations that contribute to $W_{\left(
\sigma _{i-1}\sigma _{i}\sigma _{i+1....}\right) -\left( \sigma _{j-1}\sigma
_{j}\sigma _{j+1....}\right) }$ for a given \textit{n-tuple }$\left( \sigma
_{i-1}\sigma _{i}\sigma _{i+1....}\right) -\left( \sigma _{j-1}\sigma
_{j}\sigma _{j+1....}\right) $ by fixing the relative position of the cargos 
$\xi $ and counting for all possible sequences of $0^{\prime }s$ and $%
1^{\prime }s$. We then sum over $\xi $ observing the invariance of the trace
under cyclic transformations. The results are compiled below.

\newpage

\begin{equation}
\begin{array}{llll}
(a) & \overline{W}_{(12)-(12)} & \simeq & \displaystyle\sum%
\limits_{m_{j}=1}^{M-2}\displaystyle\sum\limits_{m_{i}=1}^{M-m_{j}}\dfrac{%
(N-m_{j}-m_{i}-3)}{2}\dbinom{N-m_{j}-m_{i}-4}{M-m_{j}-m_{i}}%
x^{m_{i}}x^{m_{j}}tr(E) \\ 
&  &  &  \\ 
(b) & \overline{W}_{(121)-(12)} &  & \displaystyle\sum\limits_{m_{j}=1}^{M-2}%
\displaystyle\sum\limits_{m_{i}=1}^{M-m_{j}}\dfrac{(N-m_{j}-m_{i}-3)}{2}%
\dbinom{N-m_{j}-m_{i}-5}{M-m_{j}-m_{i}}x^{m_{i}}x^{m_{j}}tr(E) \\ 
&  &  &  \\ 
(c) & \overline{W}_{(12)-(02)} & \simeq & \displaystyle\sum%
\limits_{m_{i}=1}^{M}\dbinom{N-m_{i}-4}{M-m_{i}}\left[ \left(
N-m_{i}-3\right) x^{m_{i}}\right] tr(E) \\ 
&  &  &  \\ 
(d) & \overline{W}_{(021)-(12)} & \simeq & \displaystyle\sum%
\limits_{m_{i}=1}^{M-2}\dbinom{N-m_{i}-5}{M-m_{i}-1}\left[
(N-m_{i}-4)x^{m_{i}}\right] tr(E) \\ 
&  &  &  \\ 
(e) & \overline{W}_{(121)-(02)} & \simeq & \displaystyle\sum%
\limits_{m_{i}=1}^{M-1}\dbinom{N-m_{i}-5}{M-m_{i}-1}\left[
(N-m_{i}-4)x^{m_{i}}\right] tr(E) \\ 
&  &  &  \\ 
(f) & \overline{W}_{(021)-(02)} & \simeq & \dbinom{N-5}{M-1}\left[ (N-4)%
\right] tr(E) \\ 
&  &  &  \\ 
(g) & \overline{W}_{(02)-(02)} & \simeq & \frac{1}{2}\dbinom{N-4}{M}tr(E)%
\end{array}
\label{config}
\end{equation}
\begin{verbatim}
 
\end{verbatim}

Variables $m_{i}$ and $m_{j}$ indicate the number of \ possible consecutive
motors at the left of the cargos in each of these configurations
contributing to a given $W_{\left( \sigma _{i-1}\sigma _{i}\sigma
_{i+1....}\right) -\left( \sigma _{j-1}\sigma _{j}\sigma _{j+1....}\right) }$%
. The sums over $m_{i}$ and $m_{j}$ are estimated here in the limit of very
large systems for which $N\rightarrow \infty $ and $M\rightarrow \infty $
keeping the motor density $M/N\rightarrow \rho $ finite within the range $%
0<\rho <1.$ In this limit, the sums converge to integrals and these
integrals can be evaluated using Laplace%
\'{}%
s asymptotic method.

Consider, for instance, the sum $(a)$ in (\ref{config}). We use Stirling%
\'{}%
s formula $N!\sim \sqrt{2\pi N}N^{N}e^{-N}$ to approximate the factorials
involving the variables $N$ and $M$ \ and\ define the new variables \cite%
{marchetti}

\begin{equation}
z\equiv m_{i}/N\text{\qquad and\qquad }y\equiv m_{j}/N.  \label{z y}
\end{equation}%
$z$ and $y$ assume continuous values in this limit so that the referred sum
converges to the integral 
\begin{equation}
\overline{W}_{12-12}\sim \dfrac{N^{3}\left( 1-\rho \right) ^{4}}{2\sqrt{2\pi
N\left( 1-\rho \right) }}e^{-N\left( 1-\rho \right) \ln \left( 1-\rho
\right) }\displaystyle\int_{0}^{\rho }dy\displaystyle\int_{0}^{(\rho -y)}dz%
\sqrt{\dfrac{1-y-z}{\rho -y-z}}e^{Nh(y,z)}\dfrac{x^{Nz}x^{Ny}}{\left(
1-y-z\right) ^{3}}.  \label{w1212-1}
\end{equation}%
The function $h(y,z)$ in the expression above depends only on the sum $y+z:$

\begin{equation}
h(y,z)=h(y+z)=[1-\left( y+z\right) ]\ln [1-\left( y+z\right) ]-[\rho -\left(
y+z\right) ]\ln [\rho -\left( y+z\right) ].  \label{h}
\end{equation}%
Thus, by defining $\nu \equiv y+z,$ (\ref{w1212-1}) can be rewritten as 
\begin{equation}
\displaystyle\sum_{conf}W_{12-12}\sim \dfrac{N^{3}\left( 1-\rho \right) ^{4}%
}{2\sqrt{2\pi N\left( 1-\rho \right) }}e^{-N\left( 1-\rho \right) \ln \left(
1-\rho \right) }\displaystyle\int_{0}^{\rho }dy\displaystyle\int_{y}^{\rho
}d\nu \sqrt{\dfrac{1-\nu }{\rho -\nu }}\dfrac{e^{Nf_{2}}}{\left( 1-\nu
\right) ^{3}}  \label{somaW12-12 (3)}
\end{equation}%
where%
\begin{equation}
f_{2}=h(\nu )+\nu \ln x  \label{f1 f2 f3}
\end{equation}%
In order to apply Laplace%
\'{}%
s method for estimating the above integral, it is convenient to change the
order of the integration observing that 
\begin{equation}
\displaystyle\int_{0}^{\rho }dy\displaystyle\int_{y}^{\rho }d\nu (\cdot )%
\hspace{0.1in}=\hspace{0.1in}\displaystyle\int_{0}^{\rho }d\nu \displaystyle%
\int_{0}^{\nu }dy(\cdot )  \label{inversao na ordem}
\end{equation}%
With this change, the integral in $y$ becomes trivial and the double
integral in (\ref{somaW12-12 (3)}) reduces to 
\begin{equation}
I=\displaystyle\int_{0}^{\rho }\nu \sqrt{\dfrac{1-\nu }{\rho -\nu }}\dfrac{1%
}{\left( 1-\nu \right) ^{3}}e^{Nf_{2}(\nu )}d\nu  \label{If2}
\end{equation}%
which can be estimated by its maximum at large $N$ . For this, notice that $%
f_{2}(\nu )$ has a maximum at%
\begin{equation}
\nu _{\max }=\frac{1-x\rho }{1-x}.  \label{ni max}
\end{equation}%
Thus, if

\begin{equation}
\begin{array}{l}
\text{(A)}%
\begin{array}{l}
x\rho <1%
\end{array}%
\end{array}
\label{condicao A2}
\end{equation}%
the condition $\nu _{\max }<\rho $ can not be satisfied and the maximum
contribution to the integral in (\ref{If2}) comes from the extremum of the
interval at $\nu =0$ which is a local maximum. The result is \cite{murray}, 
\begin{equation}
I_{f_{2}}^{A,C}\sim \left[ \left( 1-\rho \right) \dfrac{N}{2}\right] \frac{1%
}{N^{2}}\frac{1}{\sqrt{\rho }}\frac{\exp [-N\rho \ln \rho ]}{\left[ \ln
(x\rho )\right] ^{2}}.  \label{If2 estimativa A e C}
\end{equation}%
If however,

\begin{equation}
\text{(B)}%
\begin{array}{l}
x\rho >1%
\end{array}
\label{condicao B2}
\end{equation}%
then $\nu _{\max }$ is localized inside the integration interval so that the
integral in (\ref{If2}) is estimated as \cite{murray} 
\begin{equation}
I_{f_{2}}^{B}\sim \left[ \left( 1-\rho \right) \dfrac{N}{2}\right] \frac{%
(x-1)^{2}}{\left( x\right) ^{3/2}}\frac{1}{(1-\rho )^{5/2}}\frac{(1-x\rho )}{%
1-x}\sqrt{\frac{2\pi }{N}}\exp \left\{ N\left[ \ln x-(1-\rho )\ln \tfrac{%
(x-1)}{(1-\rho )}\right] \right\} .  \label{If2 estimativa B}
\end{equation}

We use this same procedure to estimate all the remaining terms in expression
(\ref{<V(2)>}). We merely quote the results below, making some extra
comments when necessary.

For estimating the sum indicated as $(b)$ in (\ref{config}) we notice that
the difference $\displaystyle\sum\limits_{m_{j},m_{i}}\Delta _{m_{j},m_{i}}$
between the expression in the RHS of $(b)$ and the sum in $(a)$ is%
\begin{eqnarray}
\displaystyle\sum\limits_{m_{j},m_{i}}\Delta _{m_{j},m_{i}} &\equiv &%
\displaystyle\sum\limits_{conf}W_{121-12}-\displaystyle\sum%
\limits_{conf}W_{12-12}  \notag \\
&=&-\dfrac{(N-M)^{2}}{2}\displaystyle\sum\limits_{m_{j}=1}^{M-1}\displaystyle%
\sum\limits_{m_{i}=1}^{M-m_{j}}\dbinom{N-m_{j}-m_{i}-3}{M-m_{j}-m_{i}}\dfrac{%
x^{m_{j}}x^{m_{i}}}{N-m_{j}-m_{i}}  \label{Delta 1}
\end{eqnarray}%
Applying Laplace%
\'{}%
s method to the resulting integrals after taking the thermodynamic limit, it
gives 
\begin{eqnarray}
\displaystyle\sum\limits_{m_{j},m_{i}}\Delta _{m_{j},m_{i}} &\sim &-\left[ 
\dfrac{N^{3}(1-\rho )^{5}}{2\sqrt{2\pi N(1-\rho )}}e^{-N\left( 1-\rho
\right) \ln \left( 1-\rho \right) }\right]  \notag \\
&&\times \left\{ 
\begin{array}{l}
\dfrac{1}{N^{2}}\dfrac{1}{\sqrt{\rho }}\dfrac{\exp \left( -N\rho \ln \rho
\right) }{\left( \ln x\rho \right) ^{2}},\quad \qquad x\rho <1 \\ 
\text{or} \\ 
\dfrac{-\sqrt{\rho }}{\left( x\right) ^{5/2}}(1-x\rho )\dfrac{(1-x)^{2}}{%
(1-\rho )^{7/2}}\sqrt{\dfrac{2\pi }{N}}\exp \left\{ N\left[ \ln x-(1-\rho
)\ln \tfrac{(x-1)}{(1-\rho )}\right] \right\} ,\text{ } \\ 
\hspace{2.5in}\text{\quad\ \ \ \ \ \ \ \ }x\rho >1\ \hspace{0.1in}\text{ }%
\end{array}%
\right.
\end{eqnarray}

The sum over configurations that contribute to $W_{12-02}$ in (\ref{config}%
)\ - $(c)$ is estimated through the asymptotic behavior of a single
integral, which gives%
\begin{eqnarray}
\displaystyle\sum_{conf}W_{12-02} &\sim &\dfrac{N^{2}(1-\rho )^{4}}{\sqrt{%
2\pi N(1-\rho )}}e^{-N\left( 1-\rho \right) \ln \left( 1-\rho \right) }
\label{somaW12-02 (2)} \\
&&\times \left\{ 
\begin{array}{l}
\dfrac{1}{N}\dfrac{1}{\sqrt{\rho }}\dfrac{\exp \left( -N\rho \ln \rho
\right) }{\left\vert \ln x\rho \right\vert },\text{ \qquad\ \ \ \ \ \ \ \ \
\ \ }x\rho <1 \\ 
\text{or} \\ 
\dfrac{1}{\left( x\right) ^{3/2}}\dfrac{(1-x)^{2}}{(1-\rho )^{5/2}}\exp
\left\{ N\left[ \ln x-(1-\rho )\ln \dfrac{(x-1)}{(1-\rho )}\right] \right\} 
\sqrt{\dfrac{2\pi }{N}},\qquad \text{ }x\rho >1\ 
\end{array}%
\right.  \notag
\end{eqnarray}

Analogous procedures are used to estimate the sums over configurations of
the kind $W_{021-12}$ and $W_{121-02}$ in (\ref{config}) \ - $(d)$ and $(e),$%
which coincide in this limit:

\begin{eqnarray}
\displaystyle\sum_{conf}W_{021-12} &\sim &\displaystyle\sum_{conf}W_{121-02}%
\sim \dfrac{N^{2}(1-\rho )^{4}}{\sqrt{2\pi N(1-\rho )}}e^{-N\left( 1-\rho
\right) \ln \left( 1-\rho \right) }  \label{W021-12 fi} \\
&&\times \left\{ 
\begin{array}{l}
\dfrac{1}{N}\sqrt{\rho }\dfrac{\exp \left( -N\rho \ln \rho \right) }{%
\left\vert \ln x\rho \right\vert },\qquad \qquad \qquad \qquad \ \ \ \ \
\qquad \qquad \qquad x\rho <1 \\ 
\text{or} \\ 
\dfrac{1}{\left( x\right) ^{5/2}}\dfrac{(1-x)^{2}}{(1-\rho )^{5/2}}\exp
\left\{ N\left[ \ln x-(1-\rho )\ln \tfrac{(x-1)}{(1-\rho )}\right] \right\} 
\sqrt{\dfrac{2\pi }{N}},\quad x\rho >1%
\end{array}%
\right.  \notag
\end{eqnarray}

For the remaining sums indicated in (\ref{config})\ - $(f)$ and $(g),$ it is
sufficient to estimate the relevant contributions to $O(\sqrt{N})$ which are 
\begin{equation}
\displaystyle\sum\limits_{conf}W_{02-02}\sim \frac{N}{2}\dfrac{(1-\rho )^{4}%
}{\sqrt{2\pi N(1-\rho )\rho }}e^{-N\left[ \left( 1-\rho \right) \ln \left(
1-\rho \right) +\rho \ln \rho \right] }  \label{W02-02 comb(1)}
\end{equation}%
and

\begin{equation}
\displaystyle\sum\limits_{conf}W_{021-02}\sim N\dfrac{(1-\rho )^{4}\rho }{%
\sqrt{2\pi N(1-\rho )\rho }}e^{-N\left[ \left( 1-\rho \right) \ln \left(
1-\rho \right) +\rho \ln \rho \right] }  \label{somaW021-02 (0)}
\end{equation}

In the following, we analyze the results for the average velocity of a cargo
obtained in this limit using the estimates above.

\section{Results}

The average velocity of a cargo in the system of interacting motors and
cargos that obey the ASEP dynamics set in (\ref{dinamica}) can now be
analyzed observing the differences in the expressions obtained above for the
integrals in each of the asymptotic regions limited by the range of the
product $\rho x$ of the two variables $\rho $ and $x.$ Such differences lead
to distinct behaviors for $<v>$ characterizing different phases of the
system that, in turn, reflect the differences in the distribution of motors
along the considered microtubule.

A a fixed value of $x$ such that $x>1$ and for

\begin{equation}
\begin{array}{ll}
(a) & 0<\rho \leq 1/x%
\end{array}%
\text{ }  \label{rho variavel a}
\end{equation}%
$<v>$ is obtained from the behavior of the integrals for $\rho x\leq 1,$
resulting in%
\begin{equation}
<v>\sim \frac{(p-w)+\left\vert \ln \rho x\right\vert \left[ 2p\rho
\left\vert \ln \rho x\right\vert +(2p\rho -w)\right] -p(1-\rho )}{%
1+4\left\vert \ln \rho x\right\vert +\left[ \ln \rho x\right] ^{2}}
\label{<V(2)>(a)}
\end{equation}

Within the complementary region in which

\begin{equation}
\begin{array}{ll}
(b) & 1/x\leq \rho \leq 1%
\end{array}%
\text{ }  \label{rho variavel b}
\end{equation}%
$<v>$ is determined from the behavior of the integrals for $\rho x\geq 1;$
we find

\begin{equation}
<v>\sim -\frac{k}{x}  \label{<V(2)>(b)}
\end{equation}

Notice that for any $x<1$ the condition (a) $\rho x<1$ is always satisfied
so that the results for $<v>$ are given in this case by (\ref{<V(2)>(a)})
within the entire interval $0<\rho <1.$ Thus, for small values of $x$ the
system does not exhibits phase transitions.

The behavior of $<v>$ is shown in \emph{Fig.2} for the whole range of motor
density $\rho $, at fixed $k=1$ and $w=3$ and at various values of $x$ (\ref%
{x}).

We notice in these results that for varying $\rho $ and for $x$ slightly
above $1,$ the average velocity of the cargo changes sign. This means that
at steady state, which may be achieved at sufficiently short times after an
eventual change in motor density at the microtubule, cargos may adjust and
change their direction of propagation moving across motor clusters.

\newpage

\section{Discussions and additional remarks}

The mechanism for cargo transfer envisaged here is equivalent to a hopping
process in which the associated rates depend on site occupation. Because
motors move and their movement is affected by the presence of the cargos and
all other motors on the microtubule, the long-time dynamics of \ the system
must be examined globally.

Our results indicate that the existence of mutual interactions and the fact
that many cargos are allowed to coexist at the microtubule are determinant
for reproducing in this context the characteristics of the bidirectional
movement. We show that within a certain range of motor density a cargo in
this system executes long-range displacements in both directions. We may
argue then that long-range cargo transfer is facilitated by traffic and
specifically, by the assembly of motors into clusters, which characterizes
traffic jam. The presence of the other cargos in the system is essential for
this to occur as they function as additional obstacles that interfere in the
motor density profile. Each cargo induces aggregation of motors at its back
end. In turn, this provides the conditions for cargo to execute long-range
displacements either backwards, over the aggregate assembled at its back end
, as well as forwards, over the aggregate assembled at the back end of the
cargo in front.

As it was originally formulated the model does not account for the
possibility that a motor with one or more attached cargos may move as well.
In fact, this is the only mechanism that is usually employed to describe
cargo transport and it is the basis for \textit{coordination} or \textit{%
tug-of-war} models. Also, \ for simplicity we have considered interactions
of a cargo with a single motor at a time. Should\ a set of motors be allowed
to interact with the cargo to participate in the transfer process then the
map into the ASEP would need to be modified accordingly. We are currently
working on these possibilities by including into the dynamics (\ref{dinamica}%
) a process of the kind $20\rightarrow 02$ that recovers ergodicity of the
model \cite{goldman3}

We should emphasize that the occurrence of the long-range bidirectional
movement as a consequence of the hopping processes devised here may happen
by the action of motors of just one kind possessing a well defined polarity.
Changes in motor density and related traffic profile suffice as a mechanism
to control cargo direction and the size of the runs determined essentially
by the extent of motor clusters at jamming conditions.\ This offers a rather
straightforward explanation for the data mentioned above suggesting that the
number of long run-lengths performed by the observed beads increases
significantly as the density of motors at the microtubule increases \cite%
{beeg}. In addition, the results presented here indicate that, for
sufficiently high values of the motor density for which $\rho >1/x,$ at the
point where the model displays a phase transition, cargos would perform a
uniform movement (on average) since their velocities become independent of $%
\rho .$ Such behavior has also been observed in the same set of experiments.

As noticed by Ma and Chisholm \cite{ma}, "little is known regarding motor
traffic and how it correlates with the movement of cargo". Here, we offer a
possibility based on the idea that the transport does not require the action
of an external agent to coordinate the process, or a \textit{tug-of-war}
mechanism or even the existence of a mechanical coupling between two kinds
of motors as proposed more recently \cite{gelfand 09}. Instead, it suggests
that such coordination can be achieved by collective effects on the course
of the dynamics as the system "self-organizes" so that it presents
characteristics that reflect an internal (and global) order that does not
have its origin in the characteristics of the external medium. It is then
possible that the necessary transport in cells is accomplished just by
adjusting the density of motors at the microtubule.

Accordingly, the presence of processive motors of different polarities that
are normally required to explain the movement of a putative motor-cargo
complex would not be necessary. Transportation here is based on a mechanism
that requires formation of clusters of motors, not necessarily on their
ability to travel along long distances.

\vspace{0.3in}\vspace{0.25in}

{\LARGE Acknowledgements}

\bigskip \vspace{0.25in}

I would like to thank Domingos H.U. Marchetti for very helpful and
enthusiastic discussions regarding the conditions imposed by Fubini%
\'{}%
s theorem and the procedure used here to estimate the double integrals; and
also Elisa T. Sena for pointing to me many of the difficulties in the
initial developments of this work. I thank Scott Hines for kindly editing a
preliminary version of the text. This work had integral support from Funda%
\c{c}\~{a}o de Amparo a Pesquisa do Estado de S\~{a}o Paulo (FAPESP) -
Brazil.

\bigskip

\newpage

{\Huge Figure Caption}

\bigskip \vspace{0.2in}

\textbf{Figure 1 - }Dynamics of motors and cargos. (a) \textbf{Cargo
transfer. }It happens here through a mechanism of hopping between neighbor
motors. Due to the flexibility of the tail, the attached cargo may display
small oscillations leading to the possibility of it being caught either by
the motor at its left or by the motor at its right. The corresponding
processes 12 $\rightarrow$ 21 or 21 $\rightarrow$ 12 are represented in the
figure. (b) \textbf{The step of a motor}. The time spent by the motor with
the two heads attached to the microtubule is much larger than the time it
spends with just one of the heads attached \cite{Kawaguchi}, as a part of
the "hand-over-hand" mechanism proposed to explain the kinetics of
two-headed motor proteins \cite{howard}. Occupation of a site by a motor
occurs here whenever it is occupied by the two heads of a motor. The motor
step is then represented as $10\rightarrow 01$ which is indicated in the
figure.

\bigskip \vspace{0.2in}

\textbf{Figure 2 - }The average velocity of a cargo for $k=1$ and $w=3$ as a
function of the density $\rho $ of motors at the microtubule, for various
values of parameter $x.$

\end{document}